\documentclass[preprint,showpacs,tightenlines,preprintnumbers,amsmath,amssymb]{revtex4}
\usepackage[dvips]{graphicx}

\def\be{\begin{equation}}
\def\ee{\end{equation}}
\def\bea{\begin{eqnarray}}
\def\eea{\end{eqnarray}}

\begin{document}

\title{Spin-orbit splitting and the tensor component of 
the Skyrme interaction}

\author{G. Col\`o$^1$, H. Sagawa$^2$, S. Fracasso$^1$ 
and P.F. Bortignon$^1$}

\affiliation{$^1$ Dipartmento di Fisica, Universit\`a degli Studi and INFN,
Sezione di Milano, 20133 Milano, Italy \\
\\
$^2$ Center for Mathematical Sciences, University of Aizu,\\
Aizu-Wakamatsu, Fukushima 965-8560, Japan\\
}


\begin{abstract}
We study the role of the tensor term of the Skyrme effective 
interactions on the spin-orbit splittings in the 
N=82 isotones and Z=50 isotopes. 
The different role of the triplet-even and triplet-odd tensor 
forces is pointed out by analyzing the spin-orbit splittings 
in these nuclei. The experimental isospin dependence of these
splittings cannot be described by Hartree-Fock calculations
employing the usual Skyrme parametrizations, but is very
well accounted for when the tensor interaction is
introduced. The capability of the Skyrme forces to
reproduce binding energies and charge radii in heavy nuclei 
is not destroyed by the introduction of the tensor term.
Finally, we also discuss the effect of 
the tensor force on the centroid of the Gamow-Teller states.
\pacs{}
\end{abstract}

\maketitle

Nuclei far from the stability valley open a new test ground for nuclear 
models.  
Recently, many experimental and theoretical efforts have been
paid to study the structure and the reaction mechanisms in 
the nuclei near the drip
lines. Modern radioactive nuclear beam facilities and 
experimental detector setups have revealed several
unexpected effects in nuclei, such as the 
existence of haloes and skins~\cite{Tani},
the modifications of shell closures~\cite{Ozawa}
and the pygmy dipole resonances~\cite{Pygmy,Pygmy_Sn}.

One of the current topics 
is the role of the tensor interactions on the shell evolution 
of nuclei far from the stability line.
The role of the tensor interactions in the evolution
of the single-particle states 
was first discussed within the Skyrme Hartree-Fock (HF) 
framework by Fl. Stancu {\em et al.}, almost thirty 
years ago~\cite{Stancu}. 
However, serious attempts have never been devoted, until very recently, 
to the study of its effects on the evolution of the shell 
structure in heavy exotic nuclei. In fact, the Skyrme parameter
sets which are widely used in nuclear structure calculations do not
include the tensor contribution. This contribution was 
included only in the so-called Skyrme-Landau parametrizations
of Ref.~\cite{Liu}.

In the present paper, we discuss the isospin dependence of the shell 
structure (in particular, the spin-orbit splitting) 
of the Z=50 isotopes and N=82 isotones. We use the HF plus
Bardeen-Cooper-Schrieffer (BCS) approach, by employing 
a Skyrme parameter set plus  
the triplet-even and triplet-odd tensor zero-range 
tensor terms, which read 
\begin{eqnarray}
v_T &=& {T\over 2} \{ [ (\sigma_1\cdot k^\prime)
(\sigma_2\cdot k^\prime) - {1\over 3}(\sigma_1\cdot\sigma_2)
k^{\prime 2} ] \delta(r_1-r_2) 
\nonumber \\
&+& \delta(r_1-r_2)
[ (\sigma_1\cdot k)(\sigma_2\cdot k) - {1\over 3}
(\sigma_1\cdot\sigma_2) k^2 ] \} 
\nonumber\\
&+& U \{ (\sigma_1\cdot k^\prime) \delta(r_1-r_2) 
(\sigma_1\cdot k) - {1\over 3} (\sigma_1\cdot\sigma_2) 
[k^\prime\cdot \delta(r_2-r_2) k] \}. 
\label{eq:tensor}
\end{eqnarray}
In the above expression, 
the operator ${k}=(\nabla_1-\nabla_2)/2i$ acts on the right and
${k}'=-(\nabla_1-\nabla_2)/2i$ acts on the left. 
The coupling constants $T$ and $U$ denote the strength of the 
triplet-even and triplet-odd tensor interactions, respectively; 
we treat these coupling constants as free parameters in the following study. 
The tensor interactions (\ref{eq:tensor}) give contributions both to the 
binding energy and to the spin-orbit splitting, 
which are, respectively, quadratic and linear in the 
proton and neutron spin-orbit densities,
\begin{equation}
J_q(r)=\frac{1}{4\pi r^3}\sum_{i}v_{i}^2(2j_{i}+1)\left[j_i(j_i+1)
      -l_i(l_i+1) -\frac{3}{4}\right]R_i^2(r).
\label{eq:sd}
\end{equation}
In this expression $q=0(1)$ labels neutrons (protons), while 
where $i=n,l,j$ runs over all states having the given $q$. 
The $v_{i}^2$ is the BCS 
occupation probability of each orbital and $R_i(r)$ is 
the radial part of the wavefunction. It should be noticed that 
the exchange part of the central Skyrme interaction gives  
the same kind of contributions to the total energy density and
spin-orbit splitting. 
The central exchange and tensor contributions to the energy 
density $H$ are 
\begin{equation}
\Delta H = {1\over 2}\alpha(J_n^2+J_p^2) + \beta J_n J_p.
\label{eq:dE}
\end{equation}
The spin-orbit potential is given by
\bea
U_{s.o.}^{(q)} = {W_0\over 2r} \left( 2{d\rho_q\over dr} + 
{d\rho_{q^\prime}\over dr} \right) + \left( \alpha 
{J_q\over r} + \beta {J_{q^\prime}\over r} \right), 
\label{eq:dW}
\eea
where the first term on the r.h.s comes from the Skyrme 
spin-orbit interaction whereas the second term includes both the central
exchange and the tensor contributions, that is, $\alpha= \alpha_C +\alpha_T$ 
and $\beta=\beta_C +\beta_T$. The central exchange contributions are
written in terms of the usual Skyrme parameters, 
\begin{eqnarray}
\alpha_C & = & {1\over 8}(t_1-t_2) - {1\over 8}
(t_1x_1+t_2x_2) \nonumber\\
\beta_C & = & -{1\over 8}(t_1x_1+t_2x_2).  
\label{eq:dWc}
\end{eqnarray}
Basic definitions of all quantities derived from the Skyrme 
parameters can be found in Refs.~\cite{Skyrme_old,Chabanat}.
The tensor contributions are expressed as 
\begin{eqnarray}
\alpha_T & = & {5\over 12}U \hfill\nonumber\\
\beta_T & = & {5\over 24}(T+U).
\label{eq:dWT}
\end{eqnarray}
The central exchange contributions (\ref{eq:dWc}) have been 
neglected when fitting most of the Skyrme parameter sets, 
and when performing most of the previous HF calculations.
In this work, we employ the SLy5 parameter set~\cite{Chabanat} 
which has been fitted with the same protocol of the more
widely used SLy4 set and should consequently have similar
quality. In the case of SLy5, the central exchange 
contributions are included in the fit and we take 
them into account here.   

Except for the double-magic systems, we perform HF-BCS 
in order to take into account the pairing correlations.
Our pairing force is a zero-range, density-dependent one,
namely
\begin{equation}
V=V_0 \left( 1- \left( {\varrho \left( {\vec r_1+\vec r_2\over 2} \right)
\over \varrho_0} \right)^\gamma \right)\cdot\delta(\vec r_1-\vec r_2).
\label{ppforce}
\end{equation}
The parameters of this force
(that is, $V_0$=680 MeV$\cdot$fm$^{3}$, $\gamma$=1 and 
$\varrho_0$=0.16 fm$^{-3}$) 
have been fixed 
along the Z=50 isotopic chain in 
connection with the SLy4 set in Ref.~\cite{Fracasso}.
Therefore, we employ the same force here both for the neutron
and the proton pairing interactions in the 50-82 shell, neglecting small
readjustments which could be made to account for the Coulomb 
anti-pairing effect in the case of protons. 

Before coming to a detailed analysis of our results, let us 
mention the important general features associated with the tensor 
and the central exchange 
contributions to the 
spin-orbit splitting.  
The first point concerns the A-dependence (or isospin dependence) of the  
the first and second terms in the r.h.s. of Eq.~(\ref{eq:dW}). 
Since the Skyrme spin-orbit contributuion (proportional to $W_0$) 
gives a value of the spin-orbit splitting which is 
linear in the derivatives of the proton and neutron densities, 
the associated mass number and isospin dependence
is very moderate in heavy nuclei. On the other hand, the second term 
in Eq. (\ref{eq:dW}) depends on the spin density $J_q$ which has 
a more peculiar behavior. $J_q$ gives essentially no contribution
in the spin-saturated cases, but it increases 
linearly with the number of particles if only one of the
spin-orbit partners is filled. The sign of the $J_q$ will change 
depending upon the quantum numbers of the orbitals which are 
progressively filled: that is, the orbital with $j_{>}=l+1/2$ gives a positive
contribution to $J_q$ while the orbital with $j_{<}=l-1/2$ gives a 
negative contribution to $J_q$. This must be kept in mind to understand
the results which are discussed below. 

According to Ref.~\cite{Stancu}, the optimal parameters 
$\alpha_T$ and $\beta_T$ should be found in a triangle in the two dimensional 
($\alpha_T$, $\beta_T$) plane, lying in the quadrant of 
negative $\alpha_T$ and positive $\beta_T$. At that time, 
the force SIII was used. As already mentioned, we wish to use
here the Lyon forces which have been fitted using a more
complete protocol (including the neutron matter equation of
state) and have some better features like a more realistic
value of the incompressibilty $K_\infty$. Therefore, we have 
refitted the values of ($\alpha_T$, $\beta_T$) using the 
recent experimental data~\cite{Schiffer} for the single-particle 
states in the N=82 isotones and the Z=50 isotopes. We have not
tried to refit all the Skyrme parameters after including
the tensor terms. This can be left for future work.
However, we have checked that the binding energies and the r.m.s. 
charge radii of $^{132}$Sn ($^{208}$Pb) change, respectively, 
by +0.65\% and -0.17\% (+0.46\% and -0.11\%) when we include
the tensor force. The parameters we have chosen are
$\alpha_T$ = -170 MeV$\cdot$fm$^5$ and $\beta_T$ = 100
MeV$\cdot$fm$^5$. We should mention that for the force
SLy5, $\alpha_C$ = 80.2 MeV$\cdot$fm$^5$ and $\beta_C$ = -48.9 
MeV$\cdot$fm$^5$. The fact that we need significantly larger values of
($\alpha_T$, $\beta_T$) as compared to Ref.~\cite{Stancu}
can be adscribed to the fact that the effect of the 
central exchange terms, with ($\alpha_C$, $\beta_C$) having
opposite sign as the ($\alpha_T$, $\beta_T$) required by our fit, must
be counterbalanced. 

\begin{figure}[h]
\includegraphics[width=4.5in,height=6.0in,angle=-90]{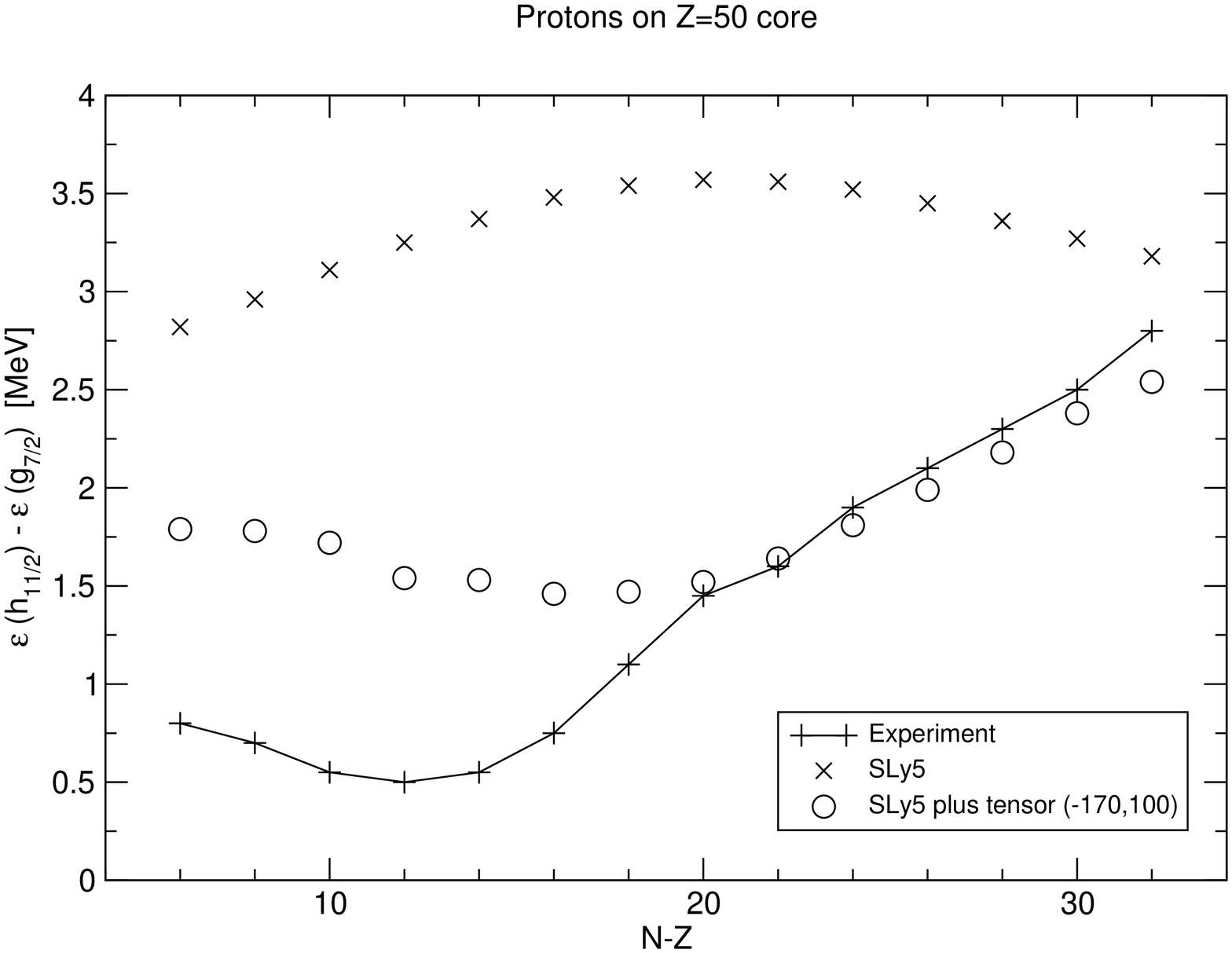}
\caption{\label{fig:Z50}
Energy differences between the 
1h$_{11/2}$ and 1g$_{7/2}$ single-proton states along the Z=50 isotopes.
The calculations are performed without (crosses) and with (circles) 
the tensor term in the spin-orbit potential (\ref{eq:dW}), on top 
of SLy5 (which includes the central exchange, or $J^2$, terms).
The experimental data are taken from ref. \cite{Schiffer}.
See the text for details.
}
\end{figure}

In Fig.~\ref{fig:Z50}, the energy differences of  
the proton single-particle states, 
$\varepsilon(h_{11/2})-\varepsilon(g_{7/2})$, along the Z=50 isotopes
are shown
as a function of the neutron excess N-Z. The original SLy5  
interaction fails to reproduce the experimental trend both qualitatively and
quantitatively. Firstly, the energy differences obtained within HF-BCS are 
much larger than the empirical data. Secondly, starting from
the double-magic $^{132}$Sn isotope, the experimental data markedly decrease 
as the neutron excess decreases and reach
about 0.5 MeV at the minimum value in $^{112}$Sn. On the other hand, 
using the HF-BCS approach with SLy5 the result is qualitatively the
opposite: the energy differences slightly increase as the neutron excess decreases, 
and there is a maximum around $^{120}$Sn. We have studied also 
several other Skyrme parameter sets and found almost the same 
trends as that of SLy5. 

In the results displayed in Fig.~\ref{fig:Z50}, 
we can see a substantial improvement due to the introduction of the 
tensor interaction with ($\alpha_T$, $\beta_T$)=(-170,100) MeV$\cdot$fm$^5$. 
This choice gives a very nice agreement with 
the experimental data in the range $20\le$ (N-Z) $\le 32$, both quantitatively 
and qualitatively. The HF+tensor results can be qualitatively understood by
simple arguments, looking at Eq.~(\ref{eq:dW}). 
In the Z=50 core, only the proton $g_{9/2}$ orbital dominates the proton spin 
density $J_p$ (cf. Eq. (\ref{eq:sd})); consequently, with a negative value 
of $\alpha_T$, the spin-orbit potential (\ref{eq:dW}) is enlarged in
absolute value (notice that $W_0$ is positive and the radial derivatives of
the densities are negative), the values 
of the proton spin-orbit splittings are increased, 
and the energy difference $\varepsilon(h_{11/2})-\varepsilon(g_{7/2})$ 
is reduced with respect to HF-BCS without tensor. This reduction is seen
better around N-Z=20: in fact, $^{120}$Sn is, to a good extent, spin-saturated
as far as the neutrons are concerned so that one gets no contribution from 
$J_n$. It should be noticed that the term in $\alpha$ does not give any 
isospin dependence to the spin-orbit potential for a fixed 
proton number, but only the term in $\beta$ can be responsible for 
the isospin dependence. In a pure HF description, 
from N-Z=6 to 14, the $g_{7/2}$ neutron orbit is gradually filled and 
$J_n$ is reduced. Then, the positive value of $\beta_T$ enlarges in absolute value 
the spin-orbit potential and increases the spin-orbit splitting, 
so that the energy difference $\varepsilon(h_{11/2})-\varepsilon(g_{7/2})$ 
becomes smaller. Because of pairing, this decrease is not so pronounced
in the results of Fig.~\ref{fig:Z50}. Moreover, from N-Z=14 to 20, the 
$s_{1/2}$ and $d_{3/2}$ neutron orbits are occupied and 
in this region the spin density is not so much changed since 
the $s_{1/2}$ orbital does not provide any contribution. Instead, 
for N-Z=20 to 32, the $h_{11/2}$ orbital is gradually filled. 
This gives a positive contribution to the spin-orbit potential (\ref{eq:dW}) 
and the the spin-orbit splitting becomes smaller. 
$\varepsilon(h_{11/2})-\varepsilon(g_{7/2})$ consequently increases, and
this effect is well pronounced in our theoretical results.
The magnitude of $\beta$ determines the slope of the isospin dependence, 
so that a larger $\beta$ would give a steeper slope. 

\begin{figure}[h]
\includegraphics[width=4.5in,height=6.0in,angle=-90]{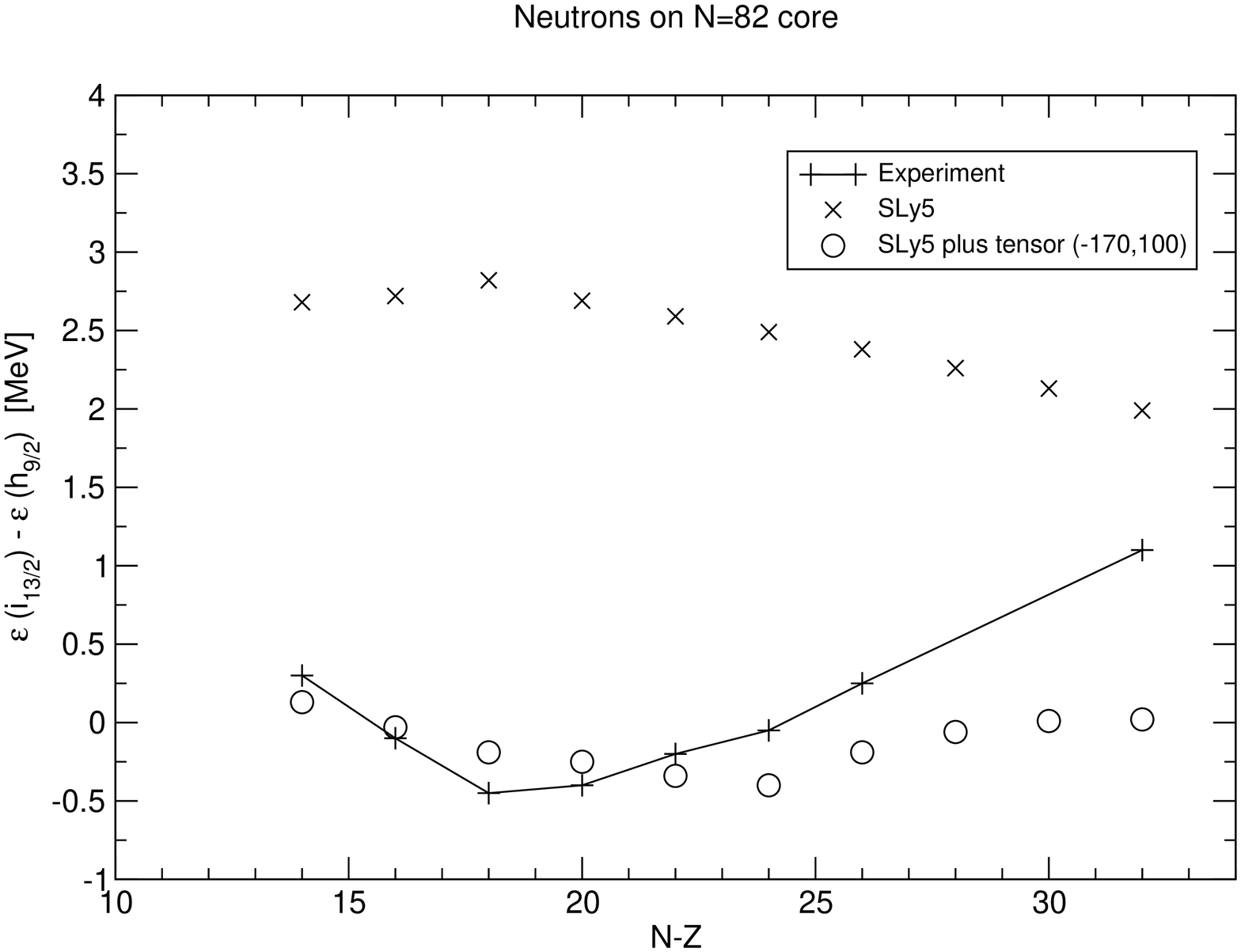}
\caption{\label{fig:N82}
Energy differences between the
1i$_{13/2}$ and 1h$_{9/2}$ single-neutron states along the N=82 isotones.
The calculations are performed without (crosses) and with (circles)
the tensor term in the spin-orbit potential (\ref{eq:dW}), on top
of SLy5 (which includes the central exchange, or $J^2$, terms).
The experimental data are taken from ref. \cite{Schiffer}.
See the text for details.
}
\end{figure}

In Fig.~\ref{fig:N82}, the energy difference $\varepsilon(i_{13/2})-\varepsilon(h_{9/2})$ 
for neutrons ouside the closed N=82 core is plotted as a function of the neutron 
excess. The notation is the same of the previous figure. 
Essentially, the same arguments already made in the previous
paragraph can be applied in order to understand the results;
simply, we should remind that the proton number is increasing
from the right (where the last nucleus displayed is $^{132}$Sn) to the left 
(where the first isotope plotted is $^{150}$Er). The 1$g_{7/2}$ and 2$d_{5/2}$ 
orbitals are rather close in energy, above the last 
occupied proton state 1$g_{9/2}$ of the Z=50 core, and their 
occupations are affected by the pairing correlations introduced by 
the BCS approximation. These two proton 
orbitals have opposite effect on the spin orbit 
potential (\ref{eq:dW}). Because of its larger value of $j$, the 
1$g_{7/2}$ orbital turns out to play a more important role 
on the spin-orbit potential, when the tensor interaction is included, in the nuclei 
with N-Z decreasing from 32 to 24. Accordingly, with positive $\beta_T$ the 
neutron spin-orbit potential is enlarged in absolute value: 
the spin-orbit splitting is made 
larger for these isotones, so that the $i_{13/2}$ orbital is pushed down and 
the $h_{9/2}$ is pushed up. These changes make the energy gap 
$\varepsilon(i_{13/2})-\varepsilon(h_{9/2})$ smaller
for the nuclei from N-Z=32 ($^{132}$Sn) to N-Z=24 ($^{140}$Ce). 
Then, the occupation of the 2$d_{5/2}$ orbital reverses the trend around
N-Z=22 ($^{142}$Nd). The theoretical trend remains the same 
until N-Z=14, since the effect of the 2$d_{3/2}$ occupation is
counterbalanced by the occupation of the 1$h_{11/2}$ which is 
not much higher and enters the active BCS space. 
  
The role of the tensor interaction due to the $\beta_T$ term 
is expected from the discussion made by J.M. Blatt and 
V.F. Weisskopf \cite{BW} for the 
deuteron. In Ref. \cite{Otsuka} the same argument was also 
presented. The role of $\alpha_T$ has not yet been examined 
in a quantitative way within the mean field calculations, as 
this term comes from the triplet-odd tensor interaction. 
The assessment of its role is new, since 
the triplet-odd tensor interaction was 
not included in the analysis of Refs. \cite{BW,Otsuka}.
Recently, Brown {\em et al.}~\cite{Brown} studied the tensor 
interactions in $^{132}$Sn and $^{114}$Sn, based on the 
parameter set Skx. They considered both positive and 
negative values of $\alpha_T$ in the HF calculations and
they concluded that $\alpha_T<$ 0 gives a better agreement 
with the experimental data. This result is consistent with 
the present systematic study of the single-particle 
states in the Z=50 isotopes and N=82 isotones, performed 
within the HF+BCS model.

The effect of the tensor interactions can be
tested on other single-particle states which are empirically
known. In this work, we have also considered the relative 
position of the 2$d_{3/2}$ and 1$h_{11/2}$ neutron states 
in $^{132}$Sn and $^{100}$Sn. In the former case ($^{132}$Sn), 
experimentally the two states are the last occupied,
with the 2$d_{3/2}$ being less bound than 1$h_{11/2}$ by 
about 240 keV (see Fig. 8 of~\cite{GL}). Theoretically, all
the mean field calculations with Skyrme or Gogny forces, as well as
the relativistic mean field (RMF) ones, result with the 
opposite ordering (see Fig. 7 of~\cite{RMP}). In particular, 
with the SLy5 force employed here, the 1$h_{11/2}$ orbital 
is less bound by 1.76 MeV. The contribution of the added 
tensor force reduces this value to 0.67 MeV. It
has to be noted that the the right position of the 2$d_{3/2}$ 
level may be rather important for the proper description of 
the low-lying dipole strength in $^{132}$Sn. In fact, the 
results obtained using a Skyrme force in Ref.~\cite{Sarchi} 
show that this strength has basically single-particle
character; but even in the calculation of Ref.~\cite{ll_rmf}, 
in which the low-lying dipole strength turns out to
present a certain degree of collectivity,  
the energy position of the levels we have mentioned is relevant
since the configurations involving the 
2$d_{3/2}$ hole contribute to 
about 50\% of the low-lying collective state. 
In the nucleus $^{100}$Sn, experimentally 
the 2$d_{3/2}$ level is more bound by 0.9 MeV (see 
Fig. 7 of~\cite{GL}). In our HF calculation with the SLy5 
force, the two levels have the right order but their
energy difference is 2.13 MeV. It is quite satisfactory that 
this difference becomes 1.33 MeV after including the 
tensor interactions.

We have already mentioned that total binding energies and
charge radii of $^{132}$Sn and $^{208}$Pb are not 
extremely sensitive to the tensor interactions. 
We have also checked in our work the effect of the tensor 
terms on the isotope shift of the Pb isotopes.
Actually, this effect is small (and does not even have 
the correct sign to reproduce experiment). 
Thus, the tensor interactions is unable to produce the
well-known empirical kink in the trend of the charge radii beyond 
$^{208}$Pb, which instead results from the introduction 
of generalized spin-orbit functionals~\cite{RF}.

Single-particle energies, and other ground-state properties, 
are not the only observables which are affected by the 
inclusion of tensor interactions. There exist excited 
states which reflect very much the behavior of the
spin-orbit splittings. One of them is the well-known
Gamow-Teller resonance (GTR). We have made a simple
estimate of the effect of the tensor interactions 
on the GTR centroid by using the sum rules. 
In charge-exchange RPA calculations,
the following sum rules are satisfied~\cite{AK},
\begin{eqnarray}
m_-(0) - m_+(0) & = & \langle 0 \vert [F^\dagger,F] \vert 0\rangle,
\nonumber\\
m_-(1) + m_+(1) & = & \langle 0 \vert [F^\dagger, [H,F]] \vert 0\rangle.
\label{sum_rules}
\end{eqnarray}
In the above expressions, $m_\pm(k)$ denotes the $k$-th 
moment of the strength in the $\Delta T_z=\pm 1$ channel: 
in particular, we are considering the non energy-weighted sum
rule $m(0)$ and the energy-weighted sum rule $m(1)$.
The associated operators $F$ and $F^\dagger$ act, respectively, 
in the $\Delta T_z=-1$ and $\Delta T_z=+1$ channels. In 
the Gamow-Teller case, they read
\begin{eqnarray}
F & = & \sum_i \vec\sigma(i)t_-(i), \nonumber\\
F^\dagger & = & \sum_i \vec\sigma(i)t_+(i).
\end{eqnarray}
Moreover, in Eq. (\ref{sum_rules}), $H$ is the total
Skyrme Hamiltonian and $\vert 0\rangle$ is the HF ground
state. In nuclei with neutron excess, the contributions
associated with the $\Delta T_z=+1$ channel are negligible
and we can approximate the GT centroid as
\begin{equation}
E_{GT} = {m_-(1)\over m_-(0)} \sim
{\langle 0 \vert [F^\dagger, [H,F]] \vert 0\rangle \over
\langle 0 \vert [F^\dagger,F] \vert 0\rangle}.
\end{equation}
The contribution from the tensor interaction to the
GT centroid is obtained by replacing the total Hamiltonian
$H$ in the previous formula, with the two-body force
of Eq. (\ref{eq:tensor}). The calculation of the 
ground-state expectation value of the 
double commutator $[F^\dagger, [v_T,F]]$ has been
worked out and it gives the contribution of the 
tensor force to the GT centroid, 
\begin{equation}
\Delta E_{GT} = {4\pi\over 9(N-Z)}
\int dr r^2 [{24\over 5}(\beta-5\alpha)J_pJ_n - 
12\alpha(J_p^2+J_n^2)].
\label{deltaE_GT}
\end{equation}

We have evaluated this latter expression, for the
Sn isotopes having N-Z larger than 20, by using
our
optimal ($\alpha_T$, $\beta_T$) values of (-170, 100). 
The
results are reported in Table~\ref{table_GT}.
The numbers are not small, but this should not be
surprising since the shifts of the single-particle
states displayed in the Figures can also be of the
order of 1-2 MeV. In fact, the positive energy shift
can be expected in $^{208}$Pb as well and can be
understood as follows. The spin-orbit densities (\ref{eq:sd})
receive contribution only from the $i_{13/2}$ orbital
(in the case of neutrons) and from the 
$h_{11/2}$ orbital (in the case of protons).
From Eq. (\ref{eq:dW}), one sees that, since
$\alpha_T$ is negative and $\vert\alpha_T\vert 
> \vert\beta_T\vert$, the net effect is an increase
of the spin-orbit splitting. Consequently, the
excitation energy of the dominant unperturbed
Gamow-Teller configuration, that is, $\nu i_{13/2}
\rightarrow \pi i_{11/2}$, is shifted upwards by
the tensor correlations. With similar arguments we
can understand the numbers reported in Table~\ref{table_GT},
as we did for the values plotted in the previous
Figures. Actually, we should remind that the analysis in terms
of the sum rules is not able to tell whether 
the peak energy is affected as much as the 
centroid $m(1)/m(0)$ since the main peak does not
exhaust, as a rule, the whole strength. 

Accurate QRPA calculations of the Gamow-Teller and
spin-dipole resonances are reported in 
Ref.~\cite{Fracasso2}. The behavior of different Skyrme 
parameter sets, without the tensor contribution, is
critically discussed. In fact, no RPA or QRPA calculations 
including the two-body tensor force are presently 
available, for any kind of vibrational mode; 
accordingly, results obtained without the tensor 
interactions should be still kept as reference 
until a global refit of the Skyrme plus tensor 
parameters is carefully accomplished. The tensor
force is not only expected to produce effect on the
Gamow-Teller states. Other vibrational
states (like the low-lying 2$^+$ which in many 
systems, once more, reflects the spin-orbit 
splitting~\cite{Peru}) 
will be certainly affected. 

A further question for future work is the role of
correlations beyond mean field. As discussed at length 
in Ref.~\cite{Mahaux}, the coupling of single-particle
states to vibrational states
has the net effect of increasing the level
density around the Fermi surface by about 30\%, by shifting occupied and
unoccupied states in opposite directions. Smaller effects are
expected for the energy differences we are considering here since
these differences involve pairs of states which are either
occupied or unoccupied. The net shift may be of the
order of few hundreds of keV as estimated from 
$^{132}$Sn~\cite{unp}.

In conclusion, the present work has shed light on the necessity
to include the tensor component in the Skyrme framework. The
first attempts in this direction were focusing on the effect
of the tensor force in magic nuclei but, as we have
stressed, if the nuclei are spin-saturated the spin-orbit 
splittings are not affected at all by the tensor force. The 
experimental mesurement of the isospin dependence of single-particle
energies has opened the possibility to fit the parameters of
the zero-range effective tensor force we are employing. 
Our results show that the introduction of the tensor force can
fairly well explain the isospin dependence of energy differences
between single-particle proton states outside the Z=50 core, and
neutron states outside the N=82 core. We have not attempted to
refit a Skyrme force by including the tensor contribution, but
we have discussed, by using the case of the Gamow-Teller 
centroids, that excited state properties will also be affected
by the tensor. An ambitious refitting program of Skyrme forces
should therefore be undertaken and deformed systems should
be considered as well~\cite{priv}. This is left as a future prospect, 
together with the role of particle-vibration coupling in this
context.

\begin{acknowledgments}
We would like to thank D. M. Brink for stimulating and enlightening 
discussions. One of us (G.C.) gratefully acknowledges the 
hospitality of the University of Aizu, where the present work 
has started. The work is supported in part by the Japanese
Ministry of Education, Culture, Sports, Science and Technology
by Grant-in-Aid for Scientific Research under
the program number (C (2)) 16540259.
\end{acknowledgments}

\newpage

\begin{table}
\caption{The effect of the tensor force on the GT centroid, evaluated
by means of Eq. (\ref{deltaE_GT}) using the SLy5 parameter set and the same
parameters of the tensor force which have been fitted on the experimental
results for the single-particle states, namely
($\alpha_T$, $\beta_T$)=(-170,100) MeV$\cdot$fm$^5$. 
See the text for a discussion of these results.\label{table_GT}}
\begin{tabular}{|c|c|}
\hline
Nucleus & $\Delta E_{GT}$ [MeV] \\
\hline
$^{120}$Sn &  1.49 \\
$^{122}$Sn &  1.55 \\
$^{124}$Sn &  1.74 \\
$^{126}$Sn &  1.99 \\
$^{128}$Sn &  2.21 \\
$^{130}$Sn &  2.48 \\
$^{132}$Sn &  2.64 \\
\hline
\end{tabular}
\end{table}

\newpage


\end{document}